# Tight-Binding "Dihedral Orbitals" Approach to Electronic Communicability in Macromolecular Chains.

Ernesto Estrada[1*] and Naomichi Hatano[2]

[1]*Complex Systems Research Group*, X-rays Unit, RIAIDT, Edificio CACTUS, University of Santiago de Compostela, 15706 Santiago de Compostela, Spain

[2]Institute of Industrial Science, University of Tokyo, Komaba 4-6-1, Meguro, Tokyo, Japan

---

\* Corresponding author. Fax: 34 981 547 077.
*E-mail address*: estrada66@yahoo.com (E. Estrada)

**Abstract**

An electronic orbital of a dihedral angle of a molecular chain is introduced. A tight-binding Hamiltonian on the basis of the dihedral orbitals is defined. This yields the Green's function between two dihedral angles of the chain. It is revealed that the Green's function, which we refer to as the electronic communicability, is useful in differentiating protein molecules of different types of conformation and secondary structure.

## 1. Introduction

The main challenge for the current post-genomic research consists of starting from the gene sequence, producing the protein, then determining its three-dimensional structure and finally extracting useful biological information about the biological role of the protein in the organism [1]. Due to the tremendous amount of structural data existing today, it is necessary to develop and use new theoretical tools to extract the maximum structural information from protein structures.

One of the most important characteristics of the three-dimensional structure of a protein is its degree of folding (DOF). The first attempt to assign a quantitative measure to DOF was carried out by Randić and Krilov [2, 3]. Balaban and Rücker [4] introduced "protochirons", and more recently, Liu and Wang [5] extended this approach by including four new kinds of 3-steps path conformations for studying DOF of protein chains. One of the present authors (EE) proposed to quantify DOF by using graph spectral theory [6-12]. This index of DOF has been very useful in structure-function studies of proteins as well as in protein secondary structure classification. All these methods are phenomenological approaches based on chemical intuition. However, it is possible to derive another index of DOF on the basis of a tight-binding Hamiltonian based on orbitals centered at the dihedral angles of a linear chain. We can thereby use a Hückel-like approach to calculate an "electronic dihedral energy" of the (protein) chain as well as an "electronic dihedral partition function" [13]. Here we extend this approach by considering the "electronic communicability" between the dihedral orbitals of a macromolecular chain by defining its electronic dihedral Green's function.

## 2. The Tight-Binding "Dihedral Orbitals" Approach

One of the present authors (EE) proposed to consider a set of orbitals situated in each dihedral angle of the chain instead of using atomic-centered orbitals as the basis of the molecular wave functions [13]. This approach is similar to the one developed by Lennard-Jones and Hall, who defined "equivalent orbitals" as orbitals centered on each bond between two atoms [14-17]. They used the bond orbitals as a basis set to give molecular orbitals (LCBO-MO) [18].

If we represent the atomic orbitals as the nodes of a chain, the bond orbitals can be represented by means of the so-called line graph of the chain. The line graph $L(G)$ is the graph in which the bonds of the chain $G$ are represented as the nodes of $L(G)$. Two nodes of $L(G)$ are adjacent if the corresponding bonds in $G$ shares an atom [19]. We can extend this approach to consider the second line graph of the molecular chain in which every node represents a bond angle. This approach is known as the iterated line graph sequence and it has been used in the study of organic molecules and macromolecules [20-22]. The corresponding orbitals located on the plane formed by the bond angle are designated as the plane orbital. Finally, what we consider here is the third line graph $L^3(G)$. The resulting orbitals are "dihedral" orbitals localized at two planes formed by two bond angles.

The molecular orbitals for the dihedral electrons in the atomic chain can be written as

$$\left|\Psi_n^D\right\rangle = \sum_{i=1}^{N} C_n^D(i)\left|\phi_i^D\right\rangle,$$  (1)

where $\phi_i^D$ is a dihedral orbital located in the $i$th dihedral angle of the chain. Equation (1) represents the basis of the linear combination of dihedral orbitals to give molecular orbitals (LCDO-MO). Then we assume that the "electronic dihedral energy" of a chain can be obtained by solving a dihedral version of the Schrödinger equation:

$$\boldsymbol{H}^D\left|\Psi^D\right\rangle = \varepsilon^D\left|\Psi^D\right\rangle,$$  (2)

where the superscript D is used to designate the dihedral angles. Substituting (1) into (2), we have

$$\sum_i C_n^D(i)\boldsymbol{H}^D\left|\phi_i^D\right\rangle = \varepsilon^D \sum_i C_n^D(i)\left|\phi_i^D\right\rangle.$$  (3)

Multiplying both sides of this expression from the left by $\left\langle\phi_j\right|$ yields the secular equation

$$\sum_i C_{ni}\left(H_{ji} - \varepsilon S_{ji}\right) = 0 , \tag{4}$$

where $H_{ji} = \left\langle \phi_j \middle| \boldsymbol{H} \middle| \phi_i \right\rangle$, $S_{ji} = \left\langle \phi_j \middle| \phi_i \right\rangle$ and we have removed, for the sake of simplicity, the superscript D in the dihedral Hamiltonian. We consider that the dihedral orbitals are orthonormal, thus $S_{ij} = \delta_{ij}$ hereafter. The nontrivial solutions of the secular equation (4) are obtained by solving the determinant equation

$$\left| \boldsymbol{H} - \varepsilon \boldsymbol{I} \right| = 0 . \tag{5}$$

We assume that the Coulomb integral $\boldsymbol{H}_{ii}$ of a dihedral orbital $\phi_i$ depends only on the angle between the two planes forming the dihedral orbital. We set the Coulomb integral in the form $H_{ii} = p - V_i q$, where the effective potential $V_i$ is some function of the $i$ th dihedral angle $\varphi_i$ of the linear chain. The constant $p$ sets the origin of the energy and the constant $q$ sets the energy scale. The resonance integral $\boldsymbol{H}_{ij}$ between dihedral orbitals $\phi_i$ and $\phi_j$ is, for the moment, assumed to be zero, unless $i$ and $j$ are adjacent dihedrals in the chain, in which case we set $H_{ij} = q$. We will argue an extension in Sec. 4, but for the moment, we obtain the Hamiltonian for a linear chain having $N$ dihedral angles in the matrix form

$$H = pI - q \begin{pmatrix} V_1 & -1 & 0 & & & \\ -1 & V_2 & \ddots & & 0 & \\ & -1 & \ddots & & & \\ & & & \ddots & & \\ & 0 & & & -1 & \\ & & & & -1 & V_N \end{pmatrix} . \tag{6}$$

This matrix is a tri-diagonal matrix $\boldsymbol{H} = \left[ H_{ij} \right]$, which means that $H_{ij} = 0$ whenever $\left| i - j \right| > 1$ [23]. Below, we will use this Hamiltonian in order to obtain the minimum dihedral energy for the most folded conformer.

The orbital energy is determined by the eigenvalues of $\boldsymbol{H}$

$$\varepsilon_j = p - q \mu_j , \tag{7}$$

where $\mu_j$ is an eigenvalue of the matrix on the r.h.s. of Eq. (6). When all dihedral orbitals are fully occupied, the total electronic "dihedral" energy (dihedral energy for brief) is given by

$$E_{dih} = \sum_{j=1}^{N} 2\varepsilon_j = Np - 2q\sum_{j=1}^{N}\mu_j \,, \tag{8}$$

where $N$ is the number of dihedral angles in the linear chain and $q < 0$. We consider that the most folded conformation permits the largest overlap between dihedral orbitals, which reduces the dihedral energy to a minimum.

From now on we set $p \equiv 0$ without loss of generality, since $p$ simply sets the origin of the energy scale. This makes the Hamiltonian

$$H = \begin{pmatrix} V_1 & -q & 0 & & & \\ -q & V_2 & \ddots & & 0 & \\ & -q & \ddots & & & \\ & & \ddots & & & \\ & 0 & & & & -q \\ & & & & -q & V_N \end{pmatrix} \tag{9}$$

and the dihedral energy [13]

$$E_{dih} = 2|q|\sum_{j=1}^{N}\mu_j \,, \tag{10}$$

where we assumed $q < 0$.

## 3. Effective potential for a dihedral angle

We now determine the functional form of the effective potential $V_i$ [13]. Our approach assumes that this potential depends only on the angle $\varphi_i$ formed between the two planes determining the dihedral angle. A natural way of selecting this function is to make it the cosine of the dihedral angle. This function satisfies our intuition that when $\varphi_i = 90°$ there is no overlapping between the plane angles forming the dihedral and $V_i$ should vanish. We also consider that for $90° < \varphi_i \le 180°$ there should be no overlapping between the dihedral angles, which means $V_i = 0$ in this region. This condition is not satisfied by the function $V_i = \cos\varphi_i$, which takes negative values for $90° < \varphi_i \le 180°$. Thus we select as the effective potential the half-cosine function in the form [13]

$$V_{ii} = \frac{1}{2}\left[1 + \text{sgn}\left(\cos\varphi_i\right)\right]\cos\varphi_i, \tag{11}$$

where $\varphi_i$ is the $i$th dihedral angle of a particular configuration of the molecular chain and $\text{sgn}(x)$ is the sign. This function is equal to cosine of the angle for $0° \le \varphi_i \le 90°$ and is zero for $90° < \varphi_i \le 180°$ as desired for our effective potential. Of course, other more sophisticated potentials may be used without modifying the significance of the current approach.

## 4. Interaction between non-adjacent dihedral angles

We now introduce an extension of the tight-binding LCDO-MO Hamiltonian (9). Depending on the three-dimensional conformation of the molecule, dihedral angles not adjacent on the linear chain may be geometrically close in the three-dimensional space. In order to take this effect into account, we introduce a coupling between non-adjacent dihedral angles. That is, we introduce a function $V_{ij}$ for the non-diagonal entries of the Hamiltonian.

Here we simply consider that the through-space jump of an electron is dependent on the geometrical (not topological) distance separating the dihedral angles. Hence we assume the following potential [24, 25]

$$V_{ij} = qe^{-k\left(d_{ij} - d_{\min}\right)}, \tag{12}$$

where $k$ is the coupling constant and $d_{\min}$ is the minimal distance between a pair of non-adjacent dihedral angles. Then the LCDO Hamiltonian is modified to

$$H = \begin{pmatrix} V_1 & -q & -qe^{-k(d_{13}-d_{\min})} & \cdots & -qe^{-k(d_{1N}-d_{\min})} \\ -q & V_2 & \ddots & \cdots & -qe^{-k(d_{2N}-d_{\min})} \\ -qe^{-k(d_{31}-d_{\min})} & -q & \ddots & \cdots & \\ & & \ddots & & \\ \vdots & \vdots & & & -q \\ -qe^{-k(d_{N1}-d_{\min})} & -qe^{-k(d_{N2}-d_{\min})} & & -q & V_N \end{pmatrix}. \tag{13}$$

In order to define the distance between two non-adjacent dihedral angles we consider the distance between the centers of gravity of both angles. In further calculations of the electronic communicability in proteins we put $k = 1$ and $d_{\min} = 0$ for simplicity. These assumptions are justified by the fact that all protein backbones have the same chemical composition and that the possible minimal separation between dihedrals in such chains is always the same.

## 5. Green's function of the dihedral chain

Once the electronic Hamiltonian is given as (13), we can define the Green's function [25]. Different approaches based on Green's function to the electron transfer in proteins have been reported [26-28]. The thermal Green's function is defined by

$$G_{ji}(\beta) = \frac{1}{Z(\beta)} \langle j | e^{-\beta H} | i \rangle = \langle j | e^{-\beta(H-F)} | i \rangle, \tag{14}$$

where the partition function is [29]

$$Z(\beta) = \mathrm{Tr}\, e^{-\beta H} \tag{15}$$

and the free energy is [29]

$$F = -\frac{1}{\beta} \log Z(\beta) \tag{16}$$

at the inverse temperature $\beta$. The Green's function describes how well the dihedral angles $i$ and $j$ are electronically connected. Hereafter we refer to the Green's function between two dihedral angles as the communicability between them, motivated by another work of the present authors in which the Green's function is identified as a graph theoretic invariant related to the number of walks linking two nodes in the graph [30].

Suppose that the Hamiltonian (13) has the eigenvalues $\{\varepsilon_\nu = |q|\mu_\nu\}$ with the eigenfunctions $\{|\psi_\nu\rangle\}$. The Green's function (14), or the communicability between the $i$ th and $j$ th dihedral angles is written in the form

$$G_{ji}(\beta) = \frac{1}{Z(\beta)} \sum_{\nu=1}^{N} \langle j|\psi_\nu\rangle\langle\psi_\nu|i\rangle e^{-\beta|q|\mu_\nu} \tag{17}$$

with the partition function

$$Z(\beta) = \sum_{\nu=1}^{N} e^{-\beta|q|\mu_\nu} . \tag{18}$$

In the low-temperature limit $\beta \to \infty$, the Green's function is reduced to

$$G_{ji}(\infty) = \langle j|\psi_1\rangle\langle\psi_1|i\rangle , \tag{19}$$

where $\nu = 1$ denotes the ground state, while in the high-temperature limit $\beta \to 0$, it is reduced to

$$G_{ji}(0) = \frac{1}{N} \sum_{\nu=1}^{N} \langle j|\psi_\nu\rangle\langle\psi_\nu|i\rangle . \tag{20}$$

Hereafter we set $|q|\beta \equiv 1$ for simplicity, since $|q|$ specifies an energy scale chosen arbitrarily.

## 6. Computational results

Here we calculate the communicability function for the human transcriptional elongation factor TFIIS, a small protein of 50 amino acids that contains a Zn(2+)-binding site. The structure of this protein, determined by complete 1H and 15N NMR [31], is used to built the LCDO Hamiltonian for the backbone chain.

We compared the result for the Hamiltonian (13) with the result for the simpler version of the Hamiltonian, Eq. (9), which ignores the terms $-e^{-k(d_{ij}-d_{\min})}$ in the expression (13). In Fig. 1a and b, we illustrate the contour plots representing the dihedral angles as the $x$ and $y$ axes and the values of the electronic communicability between the dihedrals as the $z$-axis. We then fit the data points by using the weighted least square method [32] implemented in the STATISTICA package [33].

<div align="center">**Insert Fig. 1 about here.**</div>

As can be seen in Fig. 1a and b, there are small but observable differences between the two contour plots due to the consideration of the inter-dihedral distances. There is an increase in the communicability between the dihedral angles located at the top-left corner of this plot (red contour) and decrease of the communicability at the center of the plot. This observation indicates that we need to consider the inter-dihedral distances in any further calculation of the electronic communicability in protein chains.

The electronic communicability between non-adjacent bonds can be carried out by both through-space and through-bond interactions. Consequently, the communicability between two dihedral angles separated to each other in the linear chain at a large topological distance can be influenced by their geometrical separation in the space. In order to illustrate this situation, we build the contour plot of the electronic communicability in terms of the amino acid numbers in the protein chain. In Fig. 1c, we illustrate this plot for 1TFI by considering all values of $G_{ji}$ from zero to its maximum (9.490) and in Fig. 1d, we plot only the values in the range $0 \leq G_{ji} \leq 0.01$.

In Fig. 1c, we can see that the largest communicability takes place between pairs of adjacent dihedrals (main diagonal of the plot). We also note that the dihedral angles which are close to each other in the sequence also display significant communicability. The absolutely largest communicability is observed for the terminal amino acids of the chain, only because they form two of the four strand present in this chain [31]. The large communicability between the amino acids in these regions is better observed in Fig. 1d, where the upper-right and bottom-left parts of the contour are displayed in red color. Another region with large communicability in this plot is that formed by amino acids 39-47 and amino acids 5-20. The center of this region, which has the largest communicability, is given by the interaction of amino acids 15 and 43. This is an extremely important region of the human transcriptional elongation factor TFIIS protein as it represents the binding site of the Zn(2+), which is formed by Cys40, Cys43, Cys15 and Cys12 [28]. The other two regions at the center of the contour plot which display large communicability correspond to β-turns [31].

An important part of the conformational arrangement of the dihedral angles is contained in the secondary structure of proteins [34]. Consequently, the electronic communicability should be affected by the differences in the protein secondary structure. In Fig. 2 we illustrate the contour plots of six proteins having differences in their secondary structure according to the CATH protein structure classification [35]. As can be seen in the plots a and b (mainly-α protein) the largest communicability takes place for residues which have dihedral angles centered about 60°, which corresponds to the $\psi$ and $\phi$ dihedral angles of left-handed α helices. In Fig. 2c and d (α- β protein) the largest communicability appears for the amino acids forming left-handed α-helices as well as for the regions around 240°, which correspond to β-strands. This change is more evident in the plots e) and f), (mainly- β proteins) where the largest communicability appears in the regions corresponding to β-strands and turns.

**Insert Fig. 2 about here.**

## 6. Summary

We defined the Green's function of a linear chain on the basis of dihedral orbitals. It indicates the mobility, or the "communicability" of an electron between the dihedral orbitals. We demonstrated the application of the above idea to real molecules. The electronic communicability was useful in differentiating molecules of different types of conformation and secondary structure. The current approach is not limited to the consideration of the backbone dihedrals only. It can be straightforwardly extended for considering all dihedral angles in a molecule. In such a case the number of dihedrals increases dramatically but the interpretation of electronic communicability function remains unaltered. These results can also be extended to other macromolecular chains, such as DNA.

**Acknowledgements**

EE thanks the program "Ramón y Cajal", Spain for partial financial support.

**Figure captions**

Fig. 1. Plot of the electronic communicability between dihedral angles for the human transcriptional elongation factor TFIIS (1TFI) without considering the distances in the Hamiltonian (a) and considering them (b). Electronic communicability between amino acids for the same protein considering the whole range of communicability values (c) and by considering only those values between 0 and 0.01 (d). In all cases the values of communicability are normalized.

Fig. 2. Contour plot of six proteins having differences in their secondary structures. Plots a) and b) correspond to proteins with mainly-α structures (PDB codes: 1BGC and 1RIB). Plots c) and d) correspond to proteins with α-β structures (PDB codes: 1RKR and 1HGE chain B). Plots e) and f) correspond to mainly-β structures (PDB codes: 1TFI and 1SGH). In all cases the values of communicability are normalized.

a)

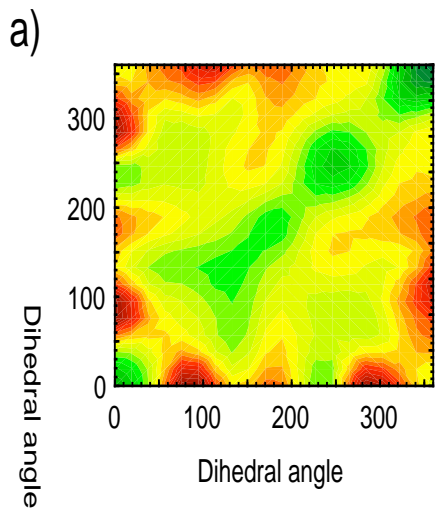

b)

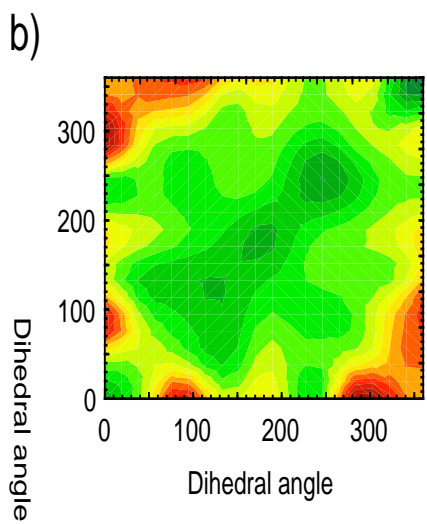

c)

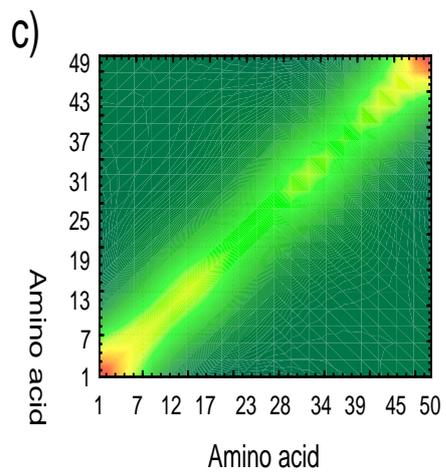

d)

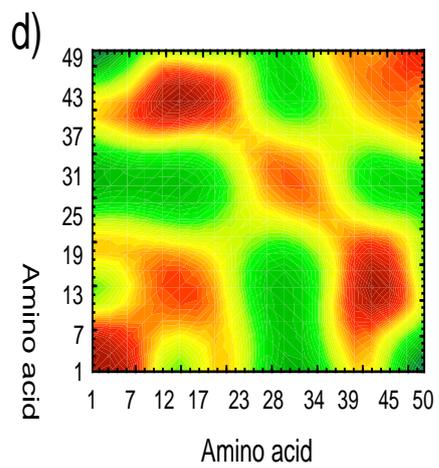

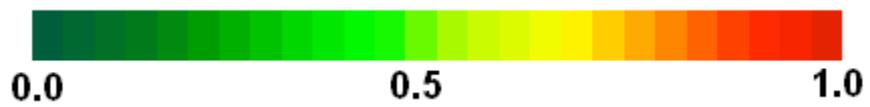

0.0          0.5          1.0

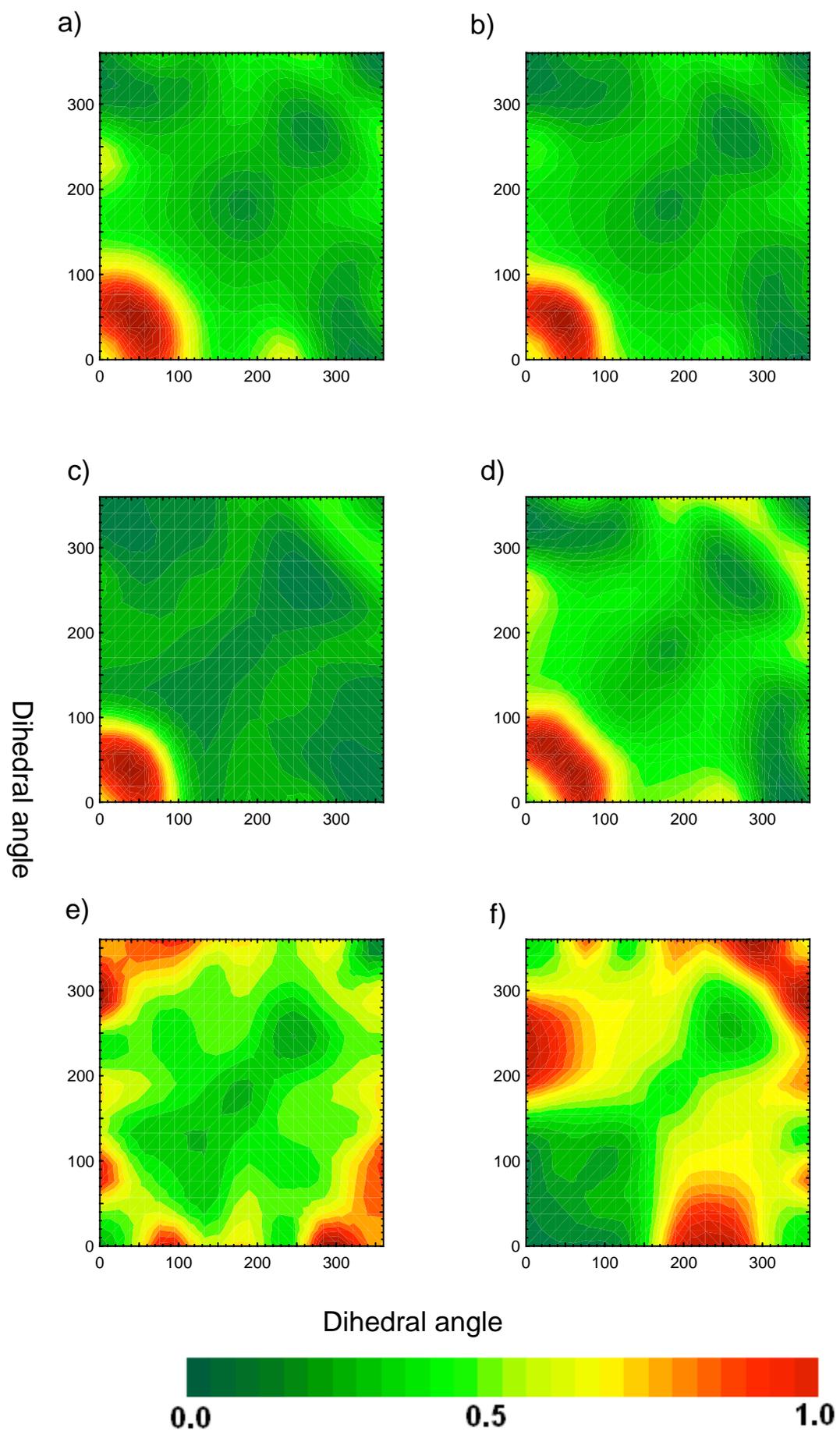